\documentclass[aps, prb, twocolumn, floatfix, showpacs]{revtex4}
\usepackage{graphicx}
\usepackage{epsfig}
\usepackage{epsf}
\usepackage{dcolumn}% Align table columns on decimal point
\usepackage{epstopdf}
\usepackage{bm}

\begin{document}

\title{Steric Quenching of the Switchable Mirror Effect}

\author{Troy C. Messina, Casey W. Miller, and John T. Markert}
\affiliation{Department  of Physics, 1 University Station - C1600,
University of Texas at Austin, Austin, TX 78712, USA}

\date{\today}% It is always \today, today,

\begin{abstract}
Scandium was substituted for yttrium to observe the effect of unit
cell size on the optical metal-to-insulator (MIT) transition in
the Y$_{1-z}$Sc$_z$H$_x$ alloy system.
The optical transmittance decreases significantly for $z\!>\!0.10$.
Simultaneous electrical resistivity measurements confirm the transition from trihydride
to dihydride behavior with increasing $z$.
These observations imply a quenching of the MIT when the unit cell volume falls below
a critical level that is consistent with the boundary between
trihydride and non-trihydride forming rare-earth elements.
A combinatoric model reveals this formation boundary corresponds to
two or more Sc per unit cell.\\
\end{abstract}
\maketitle
%: 1 intro
Yttrium and lanthanum thin films have been shown to undergo quick
and reversible transitions from a metallic-mirror to a
transparent-insulator with small changes in hydrogen
content\cite{huiberts, griessen}. The metal-to-insulator
transition (MIT) is dependent on formation of a trihydride phase
in rare-earth metals and occurs between the dihydride and
trihydride phases \cite{griessen2005}. Structural studies on La,
Y, and their alloys suggest the MIT is related to small
displacements of octahedral hydrogen, and not mediated by
structural phase transitions \cite{adolphi, kooij, remhof1,
remhof2, udovic1, udovic2, lay-prb}. This work reports on the
systematic substitution of Sc for Y to sterically control the
incorporation of hydrogen into octahedral interstices. Elemental
Sc has too small a unit cell for hydrogen to occupy octahedral
interstices, and thus cannot form a trihydride, or undergo the
metallic-mirror to transparent-insulator transition at STP. Unlike more
complex alloys, such as those with Mg and Y\cite{layh2,
shutter-effect, LaMgHx}, Sc and Y have trivalent, closed-shell
$d^1s^2$ electronic configurations. Because these metals are
chemically and structurally similar, we argue that a systematic
decrease in unit cell volume due to Sc substitution causes
Y$_{1-z}$Sc$_z$H$_x$ to cross a volume-limited
dihydride/trihydride formation boundary, where the decreasing
octahedral site radius prevents hydrogen incorporation. We show
that the optical gap seen in YH$_{3-\delta}$ is strongly
suppressed in Y$_{1-z}$Sc$_z$H$_x$ for $z\approx0.20$ and greater.
For such alloys, only dihydride-like transmittance is observed,
and resistivity measurements show increasingly metallic behavior
in the fully hydrogenated state with increasing $z$.
The proposed formation limit is discussed in terms of a simple combinatoric model.\\
%: 2 growth and characterization
\indent We deposited 100 nm thick Y$_{1-z}$Sc$_{z}$ films capped
with a 10 nm layer of palladium (Pd) on room temperature glass
substrates using electron-beam evaporation.
Base pressures lower than $10^{-8}$~torr would increase to $10^{-7}$~torr upon heating the evaporant metals.
Impurities in the vacuum presumably getter to the reactive rare earth metal deposited on the chamber walls returning the base pressure to less than $10^{-8}$~torr after some time.
All films were then deposited at $1~{\rm \AA}$/s in a vacuum better than
$10^{-8}$~torr. The Pd prevents oxidation and catalyzes hydrogen
gas (H$_{2}$) dissociation and absorption. Synthesis of the
Y$_{1-z}$Sc$_{z}$ alloys and films has been discussed previously
\cite{YScHxJAC}. The compositional dependence of unit cell lattice
parameters of as-deposited and stable dihydride films were
measured by $\theta-2\theta$ x-ray diffraction (XRD). The hcp
lattice parameters of the as-deposited films for pure Y were
$a=3.69$~\AA~ and $c=5.82$~\AA, and decreased linearly to
$a=3.31$~\AA~and $c=5.12$~\AA~for pure Sc. A similar linear
relation was found for the lattice constant of fcc dihydride
films, falling from $a=5.20$~\AA~for Y to $a=4.77$~\AA~for Sc. The
$a$ values for YH$_2$ and ScH$_2$ are within 1$\%$ of literature
values for bulk samples.
No oxide or hydroxyl impurities were observed in the XRD measurements.
The measured 177~\AA$^{3}$ unit cell
volume of the first alloy not to exhibit the MIT
(Y$_{0.80}$Sc$_{0.20}$) is approximately the unit cell volume of
Lu, the largest (and only trivalent) rare-earth element that does
not form a trihydride. Additionally, the H-H separation distance
for $z\!>\,$0.10 inferred from measured lattice parameters
becomes less than 2.1~\AA, which is reportedly the minimum
distance for H to occupy
octahedral intersticial sites\cite{westlake}.\\
%: 3 describe optical and electrical measurements
\indent At STP, the transition from yttrium dihydride to
trihydride takes seconds for films less than 500~nm thick. For our
studies, the transition was extended to hours using a
mass-flow-controlled mixture of ultra-high purity H$_2$ diluted
with Ar. The flow was directed through a modified reflection
grating optical transmission spectrometer (340 $\le \lambda \le$
960 nm). A background intensity spectrum, $I_{0}(\lambda)$, was
measured prior to each hydrogen loading, and the transmittance
calculated as
$T\left(\lambda\right)=I_{T}\left(\lambda\right)/I_{0}\left(\lambda\right)$,
where $I_{T}(\lambda)$ is the spectrum measured during hydrogen
loading. Four probe electrical resistivity measurements were made
simultaneously with the optical spectra during loading using a
1~mm $\times$ 10~mm section of the film outside of the
spectrometer beam. The resistivity of the rare-earth layer was
approximated assuming the bilayer films act as parallel resistors.
The resistivity of Pd was measured independently to be 27--29
$\mu\Omega\cdot$cm for all hydrogen concentrations. The
resistivities of pure Y and Sc were 73~$\mu\Omega\cdot$cm, and
91~$\mu\Omega\cdot$cm, respectively.\\
\begin{figure}[h]
\centering
\epsfig{file=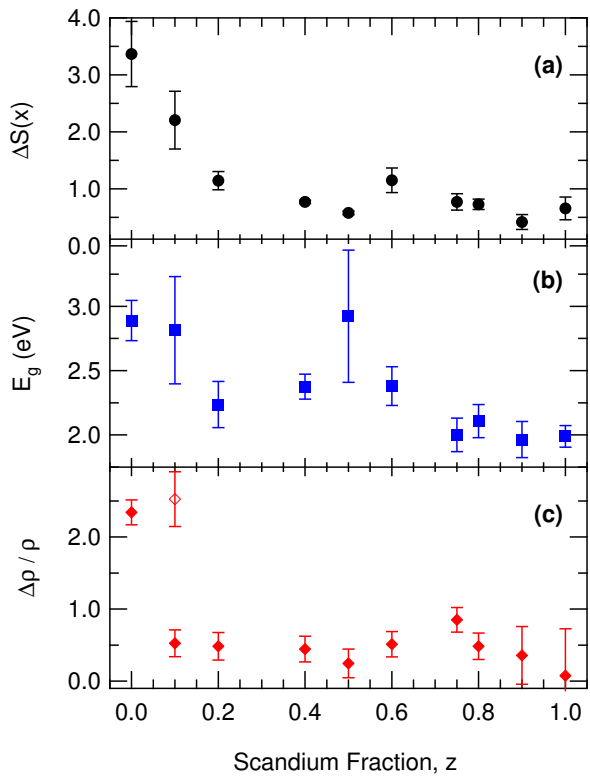,width=\linewidth}
\caption{\small{(Color online) Scandium fraction dependence in
Y$_{1-z}$Sc$_{z}$H$_x$of (a) spectral density transmittance
change, (b) best fit optical gap, and (c) relative resistivity
change. The symbol, $\diamond$, is the relative resistivity change
between as-deposited and fully hydrogenated
Y$_{0.9}$Sc$_{0.1}$H$_x$ indicating a MIT. The unloaded (presumably $x\!\sim$2) film does not show this transition suggestive of a hydrogen concentration greater than $x$=2.}} \label{spectra}
\end{figure}
%: 4 optical results show reduced MIT with z
\indent Optical transmittance measurements for $z\!\le$\,0.10 show
dramatic optical switching properties commensurate with previous
YH$_{x}$ results \cite{griessen}. The dihydride state was verified
by the characteristic transmission peak observed near
$\hbar\omega=1.8\,$eV ($\lambda\,$=\,700\,nm)
\cite{huiberts,griessen}. The hydrogen-induced transmittance
increase ($\Delta T(\lambda)$) between $x\,\approx\,$2 and
fully hydrogenated films at STP for $z\!\le0.10$ is a factor of 3 greater than that
of $z=0.20$, and a factor of 12 greater than that of $z=1.00$.
Additionally, the energy of maximum transmittance for $z\!>$\,0.10
converged to $\hbar\omega$\,=\,1.51 eV ($\lambda$\,=\,820 nm),
indicating trihydride formation is no longer observed for these alloys.
For films with $z$\,=\,0.50, the fully hydrogenated film
transmittance is trihydride-like, with a maximum at the lowest
measured energy ($\hbar\omega$\,=\,1.29 eV). This anomalous
transmittance due to phase-separated yttrium forming a trihydride
was observed for $z$~=~0.40 and $z$\,=\,0.50; phase separation was
verified by peak splitting in dihydride XRD data. The significance
of Sc substitution on the transmittance is further emphasized by
considering the spectral density change of fully hydrogenated
films relative to dihydride films for each alloy. We define the
spectral density change as
\begin{eqnarray*}
\Delta S(x) = \frac{\int T_{x}~d\lambda - \int
T_{2}~d\lambda}{\int T_{2}~d\lambda},
\end{eqnarray*}
where $T_x$ and $T_2$ are respectively the transmittance spectra
for the fully hydrogenated and dihydride films, and the
integration is performed over the full spectral width of the
measurements (620\,nm). Figure\,\ref{spectra}(a) shows a
significant reduction of the spectral density change with
increasing Sc content, implying a quenching of the MIT for $z\!>\!0.10$. \\
%: 5 LB to find gap decreases with z
\indent The Lambert-Beer law allows the optical gap to be
estimated using the absorption coefficient, $\alpha(\omega)$, in
the frequency-dependent transmittance
$T(\omega)=\,T_oexp[-\alpha(\omega)d]$, where $d$ is the film
thickness, and $T_o$ describes the system components that are
hydrogen-independent. For parabolic bands\cite{beer},
$\alpha(\omega)\,=\,C(\hbar\omega - E_g)^\nu/(k_BT)$, where $C$ is
a fit parameter and $\nu$ describes the gap in momentum
space\cite{gaps}. Fits only converged for $\nu$\,=\,2 (an allowed,
indirect gap). Fig.~\ref{spectra}(b) shows the best fit band gap
for the alloy system. $E_g$ is approximately 2.8\,eV for
$z\!<\!0.20$ (in agreement with previous results
\cite{griessen2005}), then approaches 2.0\,eV with increasing $z$
as the MIT is quenched.
The gap enhancement around $z~\approx~0.50$ is due to the aforementioned phase separation.\\
%: 6 resistivity shows reduced MIT with z
\indent Room temperature resistivity measurements made simultaneously with
optical transmittance measurements reveal a $z$-dependent
transition. This is shown in Fig.~\ref{spectra}(c), where we plot $\Delta
\rho/\rho=(\rho_{1atm} - \rho_{x=2})/\rho_{x=2}$. As the scandium content is
increased, the magnitude of the MIT decreases until the
resistivity remains at the dihydride minimum in Sc films. The
resistivity of the rare-earth layer for $z$\,=\,0.00 and
$z$\,=\,0.10 when fully hydrogenated is estimated to be 500--1000
$\mu\Omega\cdot$cm. This greater than an order of magnitude
resistivity increase is due to the formation of a trihydride. For
$z\,=\,$0.20, the change in resistivity is less than a factor
of 2, indicating a suppression in the MIT. 
%: 7 resistivity implies >2 H/metal
\begin{figure} \centering
\epsfig{file=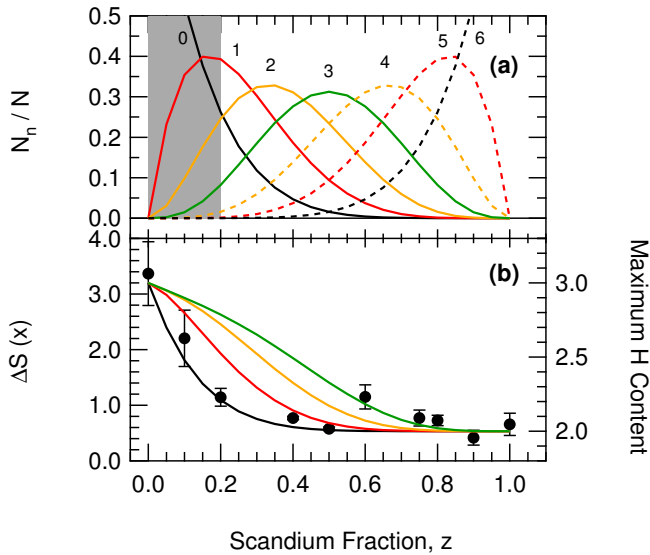,width=\linewidth}
\caption{\small{(Color online) (a) Fraction of octahedral sites
that have $n$ Sc nearest neighbors as a function of Sc fraction.
Outside of the shaded region the MIT is largely quenched. (b) The
change in spectral density transmittance scaled to the H content
predicted by restricting octahedral occupancy to sites with (dots
experimental) and model curves (from left) $n$\,=\,0, $n~\le$\,1,
$n\,\le$\,2, and $n\,\le$\,3 nearest-neighbor Sc.}}\label{maxH}
\end{figure}
Although alloys with $z\!>\,$0.10 show little or no transmittance increase beyond the dihydride concentration (suggesting that the maximum hydrogen concentration is $x\approx$\,2), the resistivity increases beyond the dihydride
minimum for all alloys with $z\,\le$\,0.90, which suggests that a
fractional amount of hydrogen is able to incorporate beyond
$x$\,=\,2 creating random scattering centers.\\
\indent Surprisingly, the z=0.1 films have greater resistivity in the unloaded
state rather than in the as-deposited state.  Unloaded films are
expected to have a dihydride composition that is characterized by a
lower resistivity than the parent metals.  This, along with the lack
of a characteristic transmittance maximum near 700 nm, may suggest the
formation of a stable super-dihydride alloy (x=2+$\delta$) for z=0.1.
Interestingly, the relative hydrogen-induced resistivity change
between the as-deposited and trihydride compositions of this alloy is
similar to that of z=0.0 (Fig1(c) open symbol, $\diamond$).\\
%: 8 combinatoric model shows MIT quenched for 4 or fewer Y nn
\indent The combinatorics can be calculated to determine the
fraction of octahedral sites that have a specific number of
nearest neighbor Sc atoms. For our samples, we have an fcc lattice
with $N$ octahedral sites each having $\xi$\,=\,6 nearest-neighbor
lattice sites. For a homogeneous Y$_{1-z}$Sc$_{z}$ alloy, the
fraction of octahedral sites with $n$ Sc nearest neighbor atoms
and $\xi-n$ nearest neighbor Y atoms is given by
\begin{eqnarray}
\frac{N_n}{N} = \frac{\xi!}{(\xi-n)! n!}(1-z)^{\xi-n}z^n.\nonumber
\end{eqnarray}
Figure~\ref{maxH}(a) shows a sharp drop in the fraction of
octahedral sites with more than five Y nearest neighbors as the Sc
content increases. Assuming it is necessary to have $i$
nearest-neighbor Sc atoms in order to permit octahedral site
occupancy, the expected total hydrogen content for each alloy can
be calculated as
\begin{eqnarray}
x~=~2~+~(1-z)\sum_{i=0}^{n}N_{i}/N,\nonumber
\end{eqnarray}
where $N_{i}/N$ is the fraction of octahedral sites with $i$
nearest-neighbors. Comparing this relation to the transmittance data reveals that the MIT occurs primarily when
octahedral interstices have five or more Y nearest neighbors
(Fig.~\ref{maxH}(b)).
As the Sc content is increased, octahedral sites have fewer Y
nearest neighbors, the unit cell volume accordingly shrinks
making it difficult for H to occupy the octahedral sites, and
the MIT is thereby quenched.\\
\begin{figure}
\centering
\epsfig{file=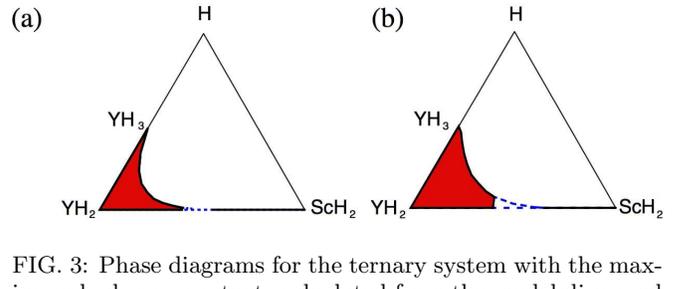, width=\linewidth}
\caption{\small{Phase diagrams for the ternary system with the
maximum hydrogen contents calculated from the model discussed in
the text. (a) Assumes $n$\,=\,0 and (b)
$n\,\le\,1$.}}\label{triangle}
\end{figure}
%: 9 phase diagram showing solid solution and phase separated regions
\indent Studies of pressure-composition isotherms for bulk Y-Sc
hydrides support our stoichiometry conclusions\cite{YScHx-old}. In
addition, the immiscibility of YH$_2$ and ScH$_2$ has been
observed and attributed to differences in atomic radii.
Thermodynamically, YH$_2$ is more stable than ScH$_2$
\cite{YScHx-old}. Hydrogen tend to reside near Y atoms, creating a
YH$_2$/ScH mixture until the dihydride filling is completed. In
alloys with near equal proportions of each metal, phase separation
is expected due to stresses induced from lattice expansion in
YH$_2$ before forming ScH$_2$. Phase diagrams generated from this
work using the maximum hydrogen concentration calculated as
discussed above are shown in Fig.~\ref{triangle}. The shaded
region represents a solid-solution phase.
The dashed lines bound the phase-separation region.
Although we have not observed a scandium trihydride in these
experiments, high pressure studies may lead to the discovery of this
material.\\
%:if we conclude n$\le$ 1, then why do we bother showing n=0?(CWM)
%: 10 summary
\indent In summary, optical transmittance spectra and electrical
resistivity measurements show that the switchable mirror effect in
the Y$_{1-z}$Sc$_{z}$H$_x$ alloy system is quenched by the loss of
available octahedral sites that results from the reduction of the
unit cell volume as Sc concentration is increased. XRD data put
this boundary at approximately the same unit cell volume as Lu and
corresponds to a previously measured volumetric limit
\cite{westlake}. Combinatoric modelling implies a significant loss
of octahedral interstices with nearest neighbor Y atoms for Sc
fractions greater than 0.10, corroborating the experimentally
observed boundary. These observations imply a natural steric
boundary for the switchable mirror effect.\\

\noindent Supported by the National Science Foundation Grant
No.~ DMR-0072365 and DMR-0605828, Robert A. Welch Foundation Grant No.~F-1191, and the Texas
Advanced Technology Program Grant No.~No. 003658-0739.

\end{document}